\documentclass[twocolumn,showpacs,preprintnumbers,amsmath,amssymb]{revtex4}
\usepackage[dvips]{graphics}
\usepackage{color}

\newcommand{\beq}{\begin{equation}}
\newcommand{\eeq}{\end{equation}}
\newcommand{\bma}{\begin{math}}
\newcommand{\ema}{\end{math}}
\newcommand{\beqa}{\begin{eqnarray}}
\newcommand{\eeqa}{\end{eqnarray}}

\def\expect#1{\langle\, #1\, \rangle}

\def\opone{\le\textbf{}\textbf{}avevmode\hbox{\small1\kern-3.8pt\normalsize1}}

\newcommand{\be}[1]{     \begin{eqnarray} \mbox{$\label{#1}$}   }

\newcommand{\ee}{\end{eqnarray}}

\newcommand\ie {{\it i.e. }}
\newcommand\eg {{\it e.g. }}

\newcommand\cf {{\it cf.  }}

\newcommand\half{\frac 1 2 }

\newcommand\ket [1] {|#1 \rangle }
\newcommand\bra [1] {\langle #1 |}

\begin{document}

\title{A Solvable Model in the Half-Filled Landau Level}

\author{Anders Karlhede}

\affiliation{Department of Physics,
Stockholm University \\ AlbaNova University Center\\ SE-106 91 Stockholm,
Sweden}

\date{\today}

\begin{abstract}
A solvable microscopic model of interacting electrons in a half-filled Landau level is presented. The Hamiltonian is a well-defined limit of the electron-electron interaction.The  low energy sector consists of a set of independent, spatially interlaced, one-dimensional systems of non-interacting fermionic dipoles, each system  with a Fermi sea ground state. The total number of dipoles in the set of Fermi seas forming the two-dimensional ground state equals the number of electrons in the system and the density of states at the Fermi level is  that of a two-dimensional Fermi gas. 
The model is obtained by generalizing, or rather adopting, the existing exact solution on a thin, but not too thin, torus to the two-dimensional bulk case. 

\end{abstract}

\pacs{73.43.Cd, 71.10.Pm, 75.10.Pq}

\maketitle

\section{Background and Result}

The low-temperature behavior of a two-dimensional gas of electrons subject to a perpendicular magnetic field has continued to be of interest ever since the discovery of the integer quantum Hall effect in 1980 \cite{vonk}, an effect that can be understood ignoring interactions between the electrons. The fractional quantum Hall effect, which is caused by these same interactions, was discovered in 1982 \cite{tsui}. The initially observed, and most prominent, fractional quantum Hall states are described by Laughlin's wave functions \cite{laugh}. Subsequently, Jain's description based on composite fermions, each consisting of an electron and an even number of quantized vortices has become an established theoretical framework—microscopically motivating many-body wave functions \cite{jain89,book}, as well as macroscopically leading to effective Chern-Simons field theories \cite{girvin,gm,zhang,read89,lopez}. The attached vortices effectively reduce the magnetic flux experienced by the composite fermions. 

The discovery by Jiang et al. in 1989 \cite{jiang} of anomalous, metal-like, transport in the half-filled lowest Landau level came as a surprise; it was a distinctive and very different behavior from what had been observed in the by then very large number of discovered quantum Hall states. In the composite fermion framework, if the magnetic field is such that the lowest Landau level is half-filled,  the picture is that all the external magnetic flux is effectively cancelled by the vortices attached to the electrons, resulting in a two-dimensional gas of non-interacting composite fermions in zero magnetic field—in agreement with the metallic behavior observed. The ground state being a Fermi sea just as for a gas of electrons. The magnetic field has disappeared in this low energy effective theory describing the system. This picture was established and made concrete by the field theory developed by Halperin, Lee and Read \cite{hlr}, and the discovery of the Fermi surface in the surface acoustic wave experiments by Willett et al.  \cite{willett90,willett98}. Furthermore, a composite fermion wave function  proposed by Rezayi and Read  was found to be in excellent agreement with exact diagonalization of small systems \cite {rr}.

Read has stressed the dipolar nature of the composite fermions, and a long-distance description of the half-filled lowest Landau level explicitly in terms of fermionic dipoles was initiated by him \cite{read94,read96,sm97,read98,sm03,sm16,ph98,dhl98,shos99,js21,milo23}.  Being neutral, these fermions decouple from the magnetic field and condense into a Fermi sea ground state.  In 2015, Son developed a field theory in terms of Dirac composite fermions that respects the particle-hole symmetry of the lowest Landau level \cite{son,sonreview}. An introduction to the Chern-Simons composite fermion theory and a comparison to the Dirac composite fermion theory, can be found in the articles by Simon \cite{simon98} and Halperin \cite{halp20}, respectively.

In this article, we pursue a microscopic approach, in terms of fermionic dipoles, to the two-dimensional gas of electrons in a half-filled Landau level. Given the success of composite fermion based theories, it is fair to ask what reasons there are for yet another approach. One reason is to try to shed light on the different, but related, long-distance approaches just mentioned. Another is that, while being a physical two-dimensional system, the quantum Hall effect is technically one-dimensional—in a single Landau level the two physical directions become conjugate in the quantum mechanical sense and hence  the system is mathematically one-dimensional. In fact, it becomes a one-dimensional lattice model with $N_e$ electrons on $N_s$ sites, where the Fock states  $\ket{n_0,n_1, n_2...n_{N_s-1}} \equiv n_0n_1 n_2...n_{N_s-1}$, with $n_i=0,1$ depending on whether site $i$ is empty or occupied by an electron, constitute a basis. The Hamiltonian describing the electron-electron interaction consists of the terms where two electrons hop symmetrically on this lattice, preserving their center of mass. Given this it is natural to ask what powerful methods available in one dimension can be applied to the quantum Hall problem. In particular, are there physically relevant models that can be solved exactly?  Given its formal one-dimensional nature this is a  case where one-dimensional methods stand a chance to be successfully applied to a physical two-dimensional system. 

One-dimensional methods of course already play acknowledged roles in the quantum Hall effect.  First of all they describe the one-dimensional edges \cite{edges}, the theory of which is intimately linked to that of the two-dimensional bulk. In high Landau levels, stripes,  whose one-dimensional edges play crucial roles, form  \cite{fks96a,fks96b,mc96,fk99}, and experiments show related anisotropy \cite{stripeexp}. In the coupled wire construction \cite{kane02,kane14,kane17,kane17b}, the quantum Hall states are obtained from interactions between one-dimensional systems. Furthermore, the one-dimensional formulation of the two-dimensional system is the technical basis for density matrix renormalization group, DMRG, calculations \cite{dmrg}.  Incidentally, the work presented here has its origin way back in a serendipitous discovery within an early effort to apply DMRG to the quantum Hall system \cite{emilDMRG,bkX}. Indications of a gapless state in the half-filled lowest Landau level on a thin cylinder were then found, leading to the exact solution on a finitely thin torus \cite{bk1,bk3}.

Still another motivation for our effort is a desire to find a microscopic understanding of the half-filled Landau level beyond the many-body wave functions, which are extremely accurate for systems with few electrons but difficult to handle beyond that. But ultimately, we were motivated by our curiosity, and limited by our abilities.

The work presented here builds on our earlier results already alluded to, where we found that,  at half-filling, on an infinitely long, but finitely thin, torus, low energy states can be explicitly mapped onto a one-dimensional theory of free fermionic  dipoles \cite{bk1,bk3}. The ground state is a Fermi sea of these dipoles and has a  large overlap with the Rezayi-Read composite fermion ansatz \cite{rr}  with an elongated Fermi sea.  This exact solution is based on: $(\it{i})$ fermionic dipole creation and annihilation operators, $(\it{ii})$ defining a restricted Hilbert space on which these fermionic operators obey  canonical anti-commutation relations, $(\it{iii})$ the Hamiltonian, on the restricted Hilbert space, being diagonalizable in terms of the dipole operators, and $(\it{iv})$ this restricted Hilbert space containing the ground state and low energy excitations.

\begin{figure}[h!]
\begin{center}
\resizebox{!}{130mm}{\includegraphics{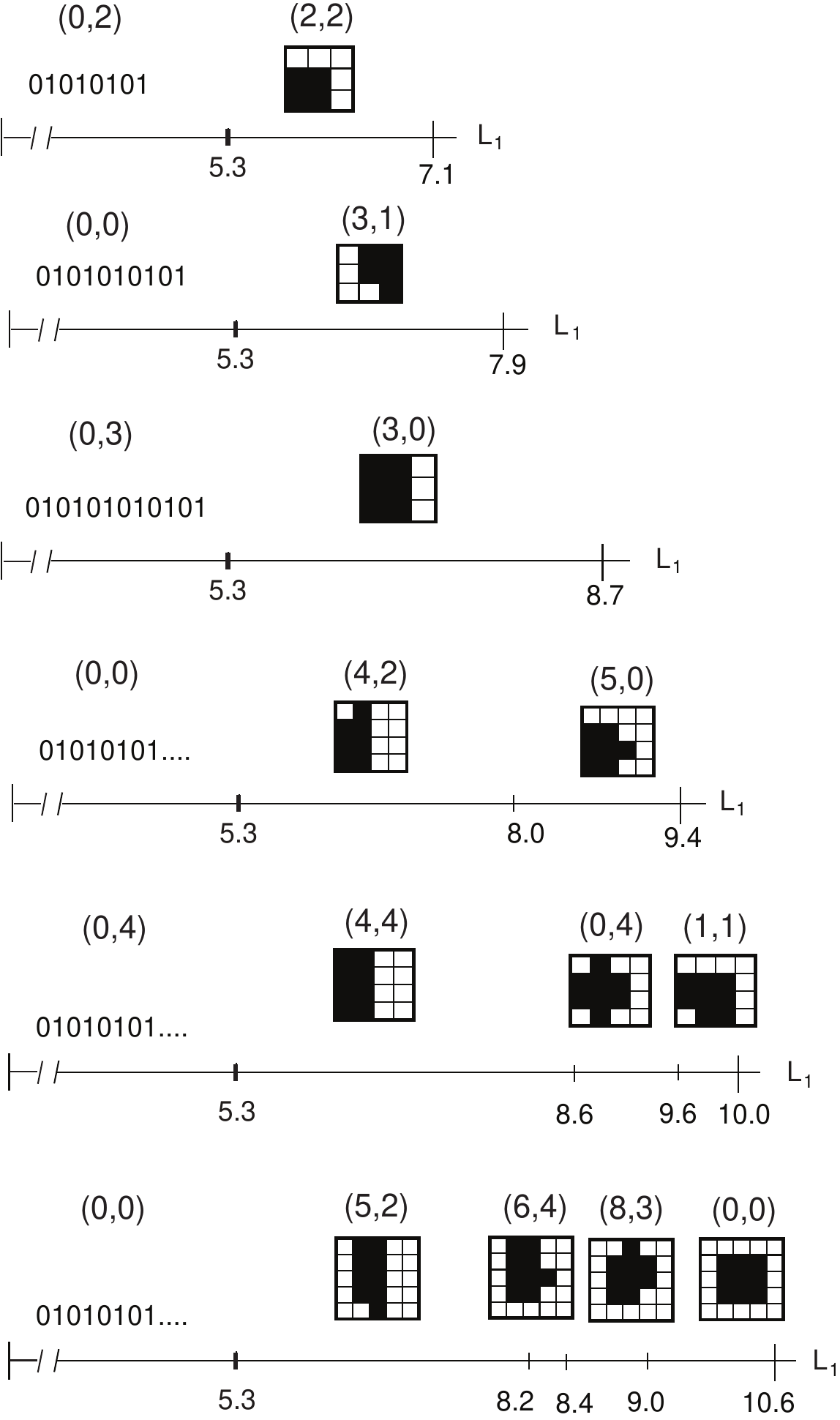}}
\end{center}\caption{\textit{{\small Ground states and corresponding quantum numbers
$(K_1,K_2)$ at $\nu=1/2$ as a function of $L_1$ for  4 to  9
      electrons (from top to bottom). 
      The results are obtained in exact
      diagonalization, using an unscreened Coulomb
      interaction. Marked $L_1$ denote transitions to new ground states.  For $L_1 < 5.3$ the ground state is  the Tao-Thouless or thin torus, TT-state, whose unit cell is 01
      when $L_1 \rightarrow 0$.  In each region $L_1 >5.3$, the state is identified as a  Rezayi-Read state with the
      displayed Fermi sea of momenta.  In the first region beyond $L_1=5.3$ the ground state is well described by the exact solution on the finitely thin torus \cite{bk1,bk3}. The largest marked $L_1$ in the figure are
      $L_1=L_2=\sqrt{4\pi N_e}$. (From \cite{bk3}.)}}}\label{phasediagrams}
\end{figure}

When the circumference of the torus increases, exact diagonalization studies indicate that the Fermi sea develops towards the standard symmetric Fermi sea of the Rezayi-Read ansatz, which is believed to describe the observed gapless state as a two-dimensional free Fermi gas of composite fermions, see Fig. \ref{phasediagrams}. It was argued that the gapless thin torus exact solution is connected to the gapless bulk state without them being separated by a phase transition. 

The above motivates the following question which is driving this investigation: Can the one-dimensional microscopic dipolar theory, which is valid on the thin torus, be extended beyond this limit? The goal would be to obtain an explicit microscopic description of the two-dimensional bulk case in terms of free fermionic dipoles, presumably described in terms of a one-dimensional theory. Here we present a solvable microscopic model with these features. 

After setting the stage in Section II, in Section III we present and motivate three conjectures that lead us to the solvable model, which we arrive at and solve in Section IV. The model is, for $N_e=nt$ electrons, based on a restricted Hilbert space with an inert stripe structure. The space is defined by its $2^{2t}$ orthonormal basis states $\prod_{i=1}^t(a_i1_{n-2}b_i0_{n-2})$, where $a_i,b_i=01 \, {\rm or} \, 10$, and  $1_{n-2} \, ,0_{n-2}$ are strings of $n-2$ inert sites that are all either occupied or empty—each such string forms an inert stripe in the two-dimensional system. For the square torus, $t=n$. The Hamiltonian is restricted to the single hopping term that acts only on the nearest neighbor interface sites $a$ and $b$: $01....10 \leftrightarrow 10....01$ in the restricted Hilbert space, and restricted static  interactions between pairs of fixed electrons. The resulting model is then nothing but the one-dimensional solvable model on the finitely thin torus described above whose ground state is a Fermi sea of neutral fermionic dipoles. This one-dimensional gapless system thus forms perpendicular to an inert background of alternating stripes of a filled and an empty Landau level, by coupling their edges together. Translational invariance of the full system gives $2n$ independent isomorphic interlaced copies of this one-dimensional system.  Hence, the total Hilbert space of the solvable model for $N_e=nt$ electrons has  $2n2^{2t}$  basis states, whereas the full Hilbert space has $\binom{2N_e} {N_e}\approx 2^{2N_e}/\sqrt{\pi N_e}$, for $N_e \gg 1$. 

The total number of dipoles in the ground state consisting of the set of Fermi seas equals the number of electrons in the system and the density of states at the Fermi level of this set  is that of a two-dimensional Fermi gas.  In general however,  the number of dipoles is not tied to the number of electrons but is determined dynamically. 
A homogeneous ground state is formed by taking a linear combination of  translated degenerate ground states. The stripes imply long range charge order but, due to the restriction on the Hilbert space, there are no gapless excitations along their edges—the gapless one-dimensional systems form across the stripes only.

In Section V we discuss the relevance of the solvable model, and comment briefly on its relation to the established theories of the half-filled lowest Landau level, as well as on higher Landau levels, stripes and wires.

\section{The Single Landau Level}\label{model}

We consider a single Landau level of $N_e$ spin-polarized electrons on a torus with lengths $L_1$ and $L_2$ in the $x$- and
$y$-directions respectively and set the magnetic length to one, $\ell_B=\sqrt{\hbar/eB}=1$. We choose one-particle states $\psi _k$, $k=0,1,\dots N_s-1$,  $N_s= L_1 L_2/2\pi$, that have
$x$-momentum $k$ and are centered at $y=-2\pi k/L_1$. The filling fraction is $\nu = N_e/N_s=p/q$.
This provides an exact mapping of the Landau level onto a one-dimensional lattice model with lattice constant $2\pi/L_1$ and $N_s$ sites.
Each site can be either empty, 0, or occupied by an electron, 1.
Thus, a basis of many particle states is provided by the set of Fock states $\{| n_0 n_1 n_2\dots n_{N_s-1} \rangle \}\equiv \{ n_0 n_1 n_2\dots n_{N_s-1} \}$, where $n_i=0,1$; alternatively,  states can be characterized by the positions (or, equivalently, the $x$-momenta) $\{k_1,k_2,...k_{N_e}\}$ of the $N_e$ electrons.

A general two-body interaction, $V(r)$,  that depends on the distance, $r$, between two electrons only, such as a Coulomb or a short-range delta function
interaction, leads to the one-dimensional Hamiltonian
\begin{eqnarray} \label{ham}
H&=& \sum_{\overset{ |m|<k \leq N_s/2} {\scriptscriptstyle{0\le j \le N_s-1}}} {\frac {V_{km} }  {1+\delta_{k,N_s/2}}} c^\dagger_{j+m}c^\dagger_{j+k}c_{j+k+m}c_j   \  \ .
\end{eqnarray}
Here $c^{\dagger}_k$ creates an electron in state $\psi_k$, $\{c_k,c_m^\dagger \}=\delta_{km}$,  $c^{\dagger}_{N_s}=c^{\dagger}_0$, and $V_{km}=V_{k,-m}$. $H$ consists of the
two-body terms that preserve the $x$-momentum, \ie the position of the center of mass of the electron pair.
$V_{km}$ is the amplitude for two electrons separated $k-m$ lattice constants to hop symmetrically to a separation of
$k+m$ lattice constants. In particular, $V_{k0}$ describes the static interaction, including the crucial exchange contribution, between two
electrons $k$ lattice constants apart. 

\begin{figure}[h!] 
\begin{center}
\resizebox{!}{45mm}{\includegraphics{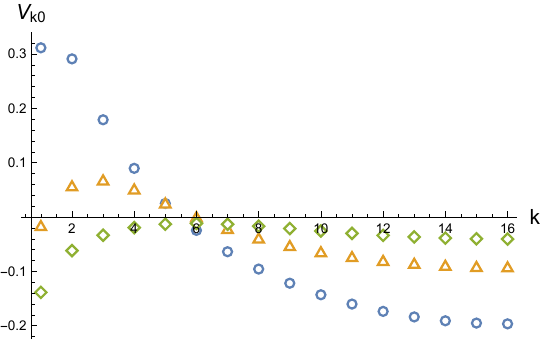}}
\end{center}\caption{\textit{{\small $V_{k0}$ for an unscreened Coulomb interaction in the lowest Landau level, and 16 electrons at $\{L_1,L_2\}=\{7.0,28.7\}$ (blue circles) and $L_1=L_2=14.18$ (orange triangles) $\{L_1,L_2\}=\{28.7,7.0\}$ (green squares); the transition to a convex $V_{k0}$ is at $L_1=5.08$. }}}\label{Vk0}
\end{figure}

\begin{figure}[h!] 
\begin{center}
\resizebox{!}{45mm}{\includegraphics{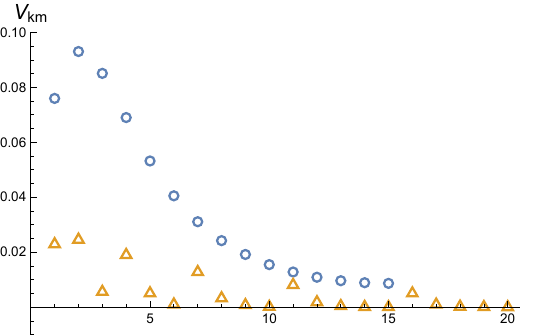}}
\end{center}\caption{\textit{{\small  $V_{k1}$ (blue circles) and  $V_{km}$, $m>1$ (orange triangles) for an unscreened Coulomb interaction in the lowest Landau level, and 16 electrons at $L_1=L_2$.  (Plotted along the $x$-axis in order  $\{k,1\}=\{2, 1\}, \{3, 1\}, ...\{16,1\}$  and $\{k,m\}=\{3,2\}, \{4,2\}, \{4,3\}, \dots \}$, respectively.) }}}\label{Vkm2}
\end{figure}

The Hamiltonian for electrons on a torus in a perpendicular magnetic field is invariant under magnetic translations $T_1$ and $T_2$. $T_1$ is related to the $x$-momentum $k$, whereas $T_2$ translate the Fock states one lattice constant. Only $T^q_2$ commutes with $T_1$, and $T_2^j$, $j=1,\dots q-1$, hence generate degenerate states from any energy eigenstate. $K_1$ and $K_2$ denote the quantum numbers associated with  $T_1$ and $T_2^q$.

\section{The TT-limit and beyond}

In the TT, Tao-Thouless or thin torus, limit, $L_1 \rightarrow 0$, $V_{km}\rightarrow 0$ if $m >0$ and the exchange contribution vanishes, making $V_{k0}$ a convex function of $k$. (From here on we restrict to $m \ge 0$, making use of $V_{k,-m}=V_{km}$.) The ground state at filling fraction $\nu=p/q$ is then determined by the purely electrostatic repulsion $V_{k0}$ and is found to be a $q$-fold degenerate ''crystal-like'' state consisting of a single Fock state $| n_0 n_1 n_2\dots n_{N_s-1} \rangle $ with a unit cell with $p$ electrons on $q$ sites, \eg $10$, $100$ and $10100$ at $\nu= 1/2, 1/3$ and $2/5$ respectively, with a gap to excitations \cite{bk3}. The domain walls between the degenerate ground states are the quasiparticles with charge $\pm e/q$. For $q$ odd these states are the abelian quantum Hall states. For  $\nu=1/3$ this is the state originally proposed by Tao and Thouless \cite{Tao83}, which thus indeed describes this observed quantum Hall effect, as they proposed. It is the exact ground state for an interaction corresponding to an extreme limit—there is no phase transition when  $L_1$ increases.  In contrast, at $\nu=1/2$ there is instead a phase transition to a gapless state. This transition occurs at $L_1 \sim  5.3 $ as indicated in Fig. \ref{phasediagrams} for different small $N_e$. 

The simple solution for quantum Hall states in the TT-limit is a consequence of $V_{k0}$ being convex and all hopping elements $V_{km}, \, m >0$, vanishing \cite{bk3}. When $L_1$ increases neither of these statements is true as the single-particle states start to overlap and, in particular,  the minimum of $V_{k0}$ seen at $k=1$ in Fig. \ref{Vk0}  develops. Thus, the long-range electron-electron repulsion is frustrated by a short-range attraction, both arising from $V_{k0}$. Such a frustration, caused by a competition between long- and short-range forces, often lead to the formation of low-dimensional structures within the system at hand \cite{stripegen}. In a half-filled high Landau level stripes form \cite{fks96a,fks96b,mc96,fk99},  but not in the half-filled lowest Landau level—this difference is caused by the Landau level dependence in $V_{km}$.  

From here on we consider, unless otherwise stated, a half-filled Landau level, and assume that $V_{km}$ are the matrix elements for the unscreened Coulomb potential in the lowest Landau level.  The $V_{k0}$ ground state, obtained by imposing  $V_{km}=0, \, m >0$, is the ''crystal-like'' gapped state obtained by placing  $N_e$ electrons, with pairwise interaction $V_{k0}$ of the type in Fig. \ref{Vk0}, on a circle with $N_s$ sites in such a way that their total interaction energy is minimized. Away from the TT-limit, when $L_1$ is not small, the problem  is how to distribute the particles making use of the minimum in $V_{k0}$ at  $k=1$ for given average distance between the electrons, which  at $\nu=1/q$ is $q$ lattice spacings. We conjecture that the solution to this combinatorial problem at $\nu=1/2$, $N_e=n^2$, $n=1,2,3,...$ and $L_1=L_2=2n{\sqrt \pi}$, is the striped Fock state with a unit cell consisting of $n$ occupied sites followed by $n$ empty sites, repeated $n$ times, \ie $\ket{(1_n0_n)_n}\equiv (1_n0_n)_n$.  Furthermore, keeping $L_1$ fixed and changing $L_2$ by changing $N_e$, we conjecture that the ground state is $(1_n0_n)_t$ at $N_e=tn$ for a range of $t$ around $t=n$: $t=n,n\pm 1, \pm 2, \dots$.  The reason for this generalization being that increasing $N_e$ while keeping $L_1$ unchanged simply changes the number of single particle states without changing the distance $2\pi/L_1$, and hence the overlap, between them. Analogous striped $V_{k0}$ ground states can be envisioned at quantum Hall fractions; at $\nu=1/3$ we propose that the $V_{k0}$ ground state is $(1_{n}0_{2n})_t$ for suitable $n,t$.

We formulate three conjectures for the ground state in the half-filled Landau level, these lead to the solvable model in the next section.

\begin{center}{\it Conjectures for $\nu=1/2$ at  $N_e=nt, \, n$ even}\end{center}

\begin{enumerate}

\item The $V_{k0}$ ground state is $\ket{(1_n0_n)_t}\equiv (1_n0_n)_t$ for $t=n$ and for other $t$ still a relevant low energy state.
\item $(1_n0_n)_t$ is the component with largest amplitude in the full $V_{km}$ ground state for $t=n$, and for other $t$ still a large component. 
\item In addition to  $V_{k0}$, it suffices to consider the single interaction component  $V_{n1}$ to arrive at a solvable model of relevance for the gapless state in the half-filled lowest Landau level.

\end{enumerate}

In addition to what has been argued above, the first two conjectures say that the Fock state with a unit cell $1_n0_n$  that minimizes the $V_{k0}$-part of the interaction is important also when the hopping terms $V_{km}, \,  m\ge 1$, are included, indicating that static and hopping terms cooperate constructively in leading to the ground state. 

We have verified the conjecture that $(1_n0_n)_t$ is the $V_{k0}$ ground state at $\nu=1/2$  for $n=4$ at $t=1,2,3,4$, \ie for $L_1=8 \sqrt \pi$ at $N_e=4,8,12,16$.  Furthermore, in these cases with full Coulomb interaction $V_{km}$, $(1_n0_n)_t$ is the ground state component with the largest amplitude. For higher $t$ we find strong evidence for the conjecture regarding the $V_{k0}$ ground state. For $L_1=L_2$ and $N_e=2^{2m}$, $m=1,2,...7$, we have verified that of all states with regular partitioning $(1_n0_n)_t$, $nt=2^{2m}$,  the one with lowest $V_{k0}$-energy is the one with $n=t$, \ie $(1_n0_n)_n$, $n=2^m$. For further details see Appendix \ref{Vk0 gs}. 

Regarding the third conjecture we consider the action of $V_{km}, m>0$, on $(1_n0_n)_t$. As illustrated in Fig. \ref{Vkm2} the dominant such matrix elements are  $V_{k1}$ and from now on we restrict to these, along with $V_{k0}$.  Of all $V_{k1}$ ($k < 3n$), only $V_{n1}$ acts on $(1_n0_n)_t$. Moreover, it acts efficiently, acting on each occurrence of both $01_n0$ and  $10_n1$ in $(1_n0_n)_t$;  it also itself acts repeatedly at one location. In addition, once $V_{n1}$ has acted, $V_{n-1,1}$ can act on the resulting state. This process continues, allowing the string of $V_{k1}$, $k=n, n-1, \dots, 2$, to act consequitively on the generated states. Thus, $(1_n0_n)_t$ is a low energy  state for $V_{k0}$, arguably the ground state, as well as a very ''hoppable'' state under  $V_{n1}$, and consequitively under $V_{k1}, \, k<n$. We interpret these facts as indicators of the state's relevance for the full $V_{km}$ ground state. We  conjecture that it is sufficient to consider $V_{n1}$ among $V_{km}, \, m > 0$, at $N_e=nt$, in order to obtain a model relevant for the half-filled lowest Landau level. 

As an aside, it is interesting to compare the action of  $V_{n1}$ on the $V_{k0}$ ground state $(1_n0_n)_t$ at $\nu=1/2$ to the corresponding action at $\nu=1/3$.  In the latter  case, the largest $V_{k1}$, \ie the one with smallest $k$ that acts on the proposed $V_{k0}$ ground state $(1_n0_{2n})_t$  is $V_{n1}$ but this acts only on $01_n0$ and not on $10_{2n}1$, action on the latter requires the considerably smaller  $V_{2n,1}$. Thus, $V_{n1}$ acts only locally at $\nu=1/3$, while at $\nu=1/2$, $V_{n1}$ acts on both subsets $01_n0$ and  $10_n1$ and consequitively globally across the different unit cells. This gives an intuitive picture of a melting of the ''crystal''  of stripes at $\nu=1/2$, leading to a gapless state, while the ''crystal''  at $\nu =1/3$ is only locally perturbed and the gap remains.

\section{The solvable model}
We here introduce, based on the conjectures above, a reduced model consisting of a restricted Hilbert space,  a correspondingly restricted Hamiltonian acting in this space and solve this model exactly under certain additional conditions. 

We consider the half-filled Landau level at $N_e=nt$, where $n\ge 4$ and is even \footnote{$n=2$ is special, there are only two Hilbert spaces ${\cal H}'_r$ and they have the  two common states $(01)_{2t}$ and $(10)_{2t}$. This is the case relevant for the finitely thin torus \cite{bk1,bk3}.}. The first step is to introduce a set of restricted isomorphic Hilbert spaces ${\cal H}'_r$, $r=0,1,\dots 2n-1$, each defined by its basis of $2^{2t}$ orthonormal Fock states $T_2^r\prod_{i=1}^t(a_i1_{n-2}b_i0_{n-2})$. Here $a_i,b_i=01 \, {\rm or} \, 10$, and  $1_{n-2} \, ,0_{n-2}$ are strings of $n-2$ sites that are all either occupied or empty,  and are inert, \ie their occupation is not affected by the restricted Hamiltonian to be defined. The symmetry operation $T_2^r$ translates the resulting state $r$ sites on the lattice, with periodic boundary conditions. The $2n$ spaces ${\cal H}'_r$ are disjoint, all the basis states are orthogonal.  The full restricted Hilbert space ${\cal H'} = \cup_{r=0}^{2n-1}{\cal H}'_r$   has $2n2^{2t}$  basis states for $N_e=nt$, whereas the full Hilbert space has
$\binom{2N_e} {N_e}\approx 2^{2N_e}/\sqrt{\pi N_e}$, for $N_e \gg 1$.

The second step is to restrict the Hamiltonian in Eq. \ref{ham} to contain, along with the static terms $V_{k0}$, only the single hopping terms $V_{n1}$ that preserve the restricted Hilbert spaces ${\cal H}'_r$, \ie those that act only on the sites $a_i,b_j$ in $T_2^r\prod_{i=1}^t(a_i1_{n-2}b_i0_{n-2})$  \footnote{Note that this restricts also the action of $V_{n1}$. A second general action of $V_{n1}$ would not preserve ${\cal H}'_r$.}. The problem then becomes identical in each restricted Hilbert space ${\cal H}'_r$ where the Hamiltonian is, dropping the index $r$,

\begin{eqnarray} \label{ham2}
H'&=& {\frac {V_{n1} }  {1+\delta_{n,N_e}}} \sum_{j=0}^{2t-1}(c^\dagger_{jn+1}c_{jn}c^\dagger_{jn+n}c_{jn+n+1} +h.c.) \nonumber \\
&+& \sum_{j=0}^{N_s-1} \sum_{ k=1}^{N_e}  {\frac {V_{k0} }  {1+\delta_{k,N_e}} }\hat n_{j+k} \hat n_j   \equiv H'_{h}+H'_{s} \, .
\end{eqnarray}
Here, $H'_{h}$ and $H'_{s}$ are the hopping and static parts  respectively.

In total, we have  $2n$ identical  Hamiltonians, each one acting in one of $2n$ isomorphic Hilbert spaces—consequently, all energy eigenstates will be $2n$-fold degenerate.  

Our final reduced model is obtained by dropping the inert sites, while taking into account that they affect the static energy via $H'_{s}$.  This gives a reduced lattice with  $\tilde N_e=2t$ electrons on $\tilde N_s=2\tilde N_e=4t$ sites, a restricted Hilbert space $\tilde{\cal H}$ defined by there being one electron on every pair $\{2j,2j+1\}$, $j=0,1,2\dots \tilde N_e -1$,  of sites. The Hamiltonian becomes
\begin{eqnarray} \label{ham5}
{\tilde H}&=& \tilde V_{21}
\sum_{j=0}^{\tilde N_e-1} [s^{+}_{j+1} s^{-}_{j}+s^{-}_{j+1}s^{+}_{j}] \nonumber \\
&+& \sum_{j=0}^{\tilde N_s-1} \sum_{ k=1}^{\tilde N_e}   \tilde V_{k0}\hat n_{j+k} \hat n_j    \, 
+  \sum_{j=0}^{\tilde N_s-1} v_j \hat n_j  + c_0 \  ,
\end{eqnarray}
where $s^{\pm}_{\tilde N_e}\equiv s^{\pm}_0$, and
\begin{equation}s^+_j=c^\dagger_{2j}c_{2j+1}\ ,  \ \ 
s^-_j=c^\dagger_{2j+1}c_{2j}\ , \ \ j=0,1\dots \tilde N_e -1 \ , \label{spinop}
\end{equation}
are spin-1/2 operators describing dipoles that are hard core bosons in $\cal{\tilde H}$, commuting on different sites but anti-commuting on the same site, see \cite{bk1,bk3}.

In Eq. \ref{ham5}, ${\tilde V}_{21} \equiv V_{n1}$ ($n<N_e$),  ${\tilde V}_{k0}$ is the $V_{k0}$-interaction between the $2t$ electrons in $\tilde{\cal H}$, $v_j$ is the $V_{k0}$-interaction of an electron at site $j$ with the inert electrons, and $c_0$ is the constant $V_{k0}$-interaction energy between the latter.  Barring the $v_j, c_0$-terms in  ${\tilde H}$ the problem  is now  identical to the one on the finitely thin torus—except for the fact that there are now $2n$ identical copies. In this limit, the reduced problem has the full symmetry, $\tilde T_1$ and  $\tilde T^2_2$, of $\tilde N_e$ electrons in a Landau level. However, the interaction $v_j$ with the inert $T_2^{2n}$-invariant background,  in general breaks the $\tilde T^2_2$-invariance to a $\tilde T^4_2$-invariance.  

The $\tilde V_{21}$-part in Eq. \ref{ham5} is the nearest neighbor Hamiltonian for the spin-1/2 XY chain. This is diagonalized for any $\tilde N_e$ by transforming the hard core bosons to fermions by a Jordan-Wigner transformation and then diagonalising the Hamiltonian by a Fourier transformation  \cite{lieb}, see also \cite{bk3}, giving 

\begin{equation} \label{hamV21}
{\tilde H}_h=2\tilde V_{21}\sum_{k}\cos{k} \ \tilde{d}^\dagger_{k} \tilde{d}_{k} \ ,
\end{equation}
where, 
\begin{eqnarray}\label{dipoles}
 \tilde{d}_{k}&=&\frac{1}{\sqrt{\tilde N_e}}\sum_{j=0}^{\tilde N_e-1} e^{-ijk}K_j s^-_j \,  , \ \ \{\tilde d^\dagger_k, \tilde d_{k'}\}=\delta_{kk'} \ ,
\end{eqnarray}
are fermionic dipole operators.
Here, $K_j=e^{i\pi
\sum_{t=0}^{j-1}s^+_t s^-_t} $ is the Klein factor and,
for states with an odd number of fermions, 
\begin{eqnarray}\label{kvalues}
k \frac {\tilde N_e} {2\pi}  &=& 0,\pm 1,\pm 2, \dots \pm (\frac {\tilde N_e}  2 -1), \frac {\tilde N_e} 2 \ ,
\end{eqnarray}
whereas for states with an even number of fermions, $k$ is shifted by $-\pi/ \tilde N_e$.
The dipoles form a  Fermi sea ground state by filling all the negative energy states  
\begin{equation} \label{gs}
\ket{GS}=\prod_{|k|>\frac \pi 2}
  \tilde{d}^\dagger_{k}\ket{010101\ldots}\ ,
\end{equation}
and give low energy excitations out of this state, which is gapless when $\tilde N_e \rightarrow \infty$.

Considering the $V_{k0}$-dependent terms in Eq. \ref{ham5},  one finds that all states in the restricted Hilbert space have the same $V_{k0}$-energy if  $V_{k0}=0$ for 
\begin{eqnarray} \label{cond}
 k=1,jn, jn \pm 1, tn-1,tn \ {\rm for} \, j=1,2\dots t-1 \, .
\end{eqnarray}
For $N_e$ large, these are few conditions. The exact diagonalization just provided holds for Eq. \ref{ham5} and the energies are, up to a common constant, independent of the values of all $V_{k0}$ not set to vanish by the given conditions. This is the {\it solvable model}.  \footnote{Obviously, $V_{n1} \rightarrow 0$ when $n\sim \sqrt{\tilde N_e} \rightarrow \infty$. However,  $\tilde V_{21}$ is just the overall scale in the Hamiltonian, which in the solvable model is assumed to be an undetermined constant rather than $V_{n1}$.}

Homogenous ground states in the full Hilbert space are obtained by forming $T_2^2$-invariant states out of the $2n$ degenerate ground states obtained from the solvable model. The homogeneity is a consequence of the ground state of the exact solution in $\tilde {\cal H}$ being homogeneous, and the state in the full Hilbert space differing simply by the introduced inert sites, see Appendix \ref{gshomo}. Although being homogeneous, the ground state has long range charge order, caused by the inert background. This is a consequence of the restricted Hilbert space.

A crucial feature of the observed gapless system in the half-filled Landau level is that it behaves as a two-dimensional Fermi gas, with its characteristic energy independent density of states. The ground state in our solvable model at $N_e=n^2$, $L_1=L_2=2n\sqrt{\pi}$, for one of the $2n$ one-dimensional systems with $\tilde N_e=2n$ electrons is the one-dimensional Fermi sea in Eq. \ref{gs}. At each of its two Fermi points there are, according to Eq. \ref{hamV21},  $2V_{21}/(2\pi /L_1)$ states within $\Delta E= 2V_{21}$. The density of states, which is the total number of states at the $4n$ Fermi points per energy interval and area, then becomes $4n/[(2\pi /L_1)L_1^2]=\pi^{-3/2}$—indeed a constant, which equals $2\rho/\sqrt{\pi}$, where $\rho$ is the constant density of states in a Landau level. 

The composite fermion approaches attach two quantized vortices to each electron in the system which then all form the two-dimensional Fermi sea. The fact that there is one composite fermion per electron determines the Fermi wave vector  to be $k_F=\sqrt{4 \pi \rho}$, in agreement with experiments \cite{willett90,willett98,hlr}.  To obtain the number of dipoles in the ground state of the exact solution we consider $N_e=n^2$ at $L_1=L_2$. A $T_2^2$-invariant ground state is formed within $\cup_{j=0}^{n-1}{\cal H}'_{2j}$.   Each pair of sites $a_i$ or $b_i$ in  $T_2^r\prod_{i=1}^n(a_i1_{n-2}b_i0_{n-2})$ is one dipole site and can be either empty or occupied. Thus there are $2n$ dipole sites in each ${\cal H}'_r$, and in total $2n^2$ dipole sites in 
$ \cup_{j=0}^{n-1}{\cal H}'_{2j}$. In the exact solution ground state, Eq. \ref{gs}, half of the dipole sites are filled.  Hence there are in total $n^2$ dipoles in the ground state Fermi sea, \ie one dipole for each electron in the system. This is the same counting as for composite fermions, but is for the dipoles determined dynamically; dipoles can be created or destroyed, not just moved out of the Fermi sea.

\section{Relevance of the solvable model}

In a solvable microscopic model obtained by restricting the Hilbert space as well as the Hamiltonian, the half-filled Landau level becomes a set of independent systems of free one-dimensional  fermionic dipoles, each set forming a Fermi sea ground state with gapless excitations.  

We here discuss the relevance of the solvable model, and its accompanying exact solution. First, we note that the model is a well-defined limit of the full problem of interacting electrons in a half-filled Landau level. Furthermore, the solution obtained is valid in the two-dimensional bulk limit, $L_1\sim L_2\rightarrow \infty$, and has  qualitative features expected of the half-filled lowest Landau level—its ground state is a Fermi sea consisting of a set of independent one-dimensional Fermi seas, and the excitations are free neutral fermions, dipoles, and hence do not couple to the magnetic field. Moreover, the density of states is that of a two-dimensional electron gas, and the set of Fermi seas contains one dipole for each electron in the system, giving the same counting as for composite fermions which in that case leads to a Fermi wave vector in agreement with experiments. 

Three conjectures, motivated by the electron-electron interaction $V_{km}$, led to the solvable model. The case of eight electrons is an illustrative example and is the smallest system that allows a non-trivial application of the mapping onto the restricted Hilbert space. Considering this system at $L_1=2L_2$ \footnote{$L_1$, which gives the lattice spacing, is chosen as its value for the square torus at $N_e=16$,  \cf Appendix \ref{Vk0 gs}.}, the exact solution ground state provided by  the solvable model consists of the four largest, and none other, of the 103 components in the full Coulomb ground state \footnote{The ground state for $\tilde N_e=4$ consists of two $\tilde T_2^2$-multiplets, each of which gives {\it two}  $T_2^8$-multiplets in each ${\cal H}_r'$. Upon repeated action by  $T_2^2$, these multiplets become  $T_2^2$-multiplets in ${\cal H'}$.}, and has overlap 0.973732 with these four components. (The probability for the four largest components in the full Coulomb state is 0.452818.) We suggest that the identification of the dominant components remains correct as $L_1\sim L_2 \rightarrow \infty$. 

The ground state obtained in exact diagonalisation depends on $N_e$, $L_1/L_2$ as well as on the interaction parameters $V_{km}$, whereas the exact solution is uniquely specified by the two discrete parameters $n$ and $t$ in  $\prod_{i=1}^t(a_i1_{n-2}b_i0_{n-2})$, defining the restricted Hilbert space.  The role of the electron-electron interaction $V_{km}$ in the exact solution is simply, but of course crucially, to identify the restricted Hilbert space and provide the hopping term $\tilde V_{21}$. In the solvable model based on that Hilbert space $\tilde V_{21}$ enters the Hamiltonian only as the overall parameter  setting the energy scale. In fact, with the prediction, in the lowest Landau level, that $t=n$ when $N_e =n^2 \rightarrow \infty$ for $L_1=L_2$, there is no free parameter in the solvable model, apart from $n=\sqrt {N_e}$. However, in higher Landau levels, changes in $V_{k0}$  presumably lead to a different relation between $n$ and $t$, as indicated by the width of the stripes where these form \cite{fks96a,fks96b,mc96,fk99,stripeexp}.

The half-filled lowest Landau level has been studied intensely, experimentally as well as theoretically, since the discovery of its distinctive experimental features in 1989 \cite{jiang}. The approach presented here should be compared to these  efforts. Different, but related, theoretical approaches exist. There is the early Chern-Simons based field theory  due to Halperin, Lee and Read \cite{hlr}, and the composite fermion wave function ansatz due to Rezayi and Read \cite{rr} for the ground state. Both of these are in the spirit of Jain's original construction of composite fermion wave functions for the fractional quantum Hall states \cite{jain89,book}, in the sense that electrons attach quantized vortices to form  composite fermions. The field theory is similar in spirit to the Chern-Simons field theory description of the quantum Hall Laughlin states \cite{girvin,gm,zhang,read89,lopez}. Then there is the related effective theory approach, initiated by Read, explicitly in terms of fermionic dipoles \cite{read94,read96,sm97,read98,sm03,sm16,ph98,dhl98,shos99,js21,milo23}. There is the more recent field theory description by Son in terms of Dirac composite fermions rather than the standard composite fermions \cite{son,sonreview}.  On the face of it these theories look quite different, Son's theory is manifestly particle-hole symmetric whereas the one by Halperin, Lee and Read is not. Moreover, in Son's theory the number of Dirac composite fermions is not tied to the number of electrons,  which is the case in the standard composite fermion approaches, microscopic as well as effective theory ones.  In spite of these differences, the theories to a large extent give the same predictions \cite{halp20}. Strong experimental evidence, as well as small system microscopic studies, support the approaches.  

The solvable microscopic model presented here, being formulated in the lowest Landau level, is manifestly particle-hole symmetric.  It is formulated explicitly in terms of dipoles, the number of which is determined dynamically and not tied to the number of electrons. Dipoles can be added or subtracted individually creating excitations that are not just particle-hole pairs relative to the Fermi surface.  However, in the ground state of the solvable model there is one dipole per electron which is the same counting as for composite fermions. These facts agree with Son's theory in terms of Dirac composite fermions.

The restricted Hilbert space, with its inert sites, implies long range charge order in the form of stripes in the solvable model. It should be noted however, that the would be gapless quantum Hall edge states along the stripes are excluded by the restriction on the Hilbert space—the gapless low energy states in the exact solution are across the stripes only. Nevertheless, the long range order is  unphysical in the lowest Landau level, while such an order is observed in higher Landau levels \cite{stripeexp,fks96a,fks96b,mc96,fk99}.  The solvable model is a well-defined limit of the full problem in a half-filled Landau level. Consequently, when ignored lowest Landau level interactions are turned on, in the full Hilbert space, the stripes are expected to melt and the long range order to disappear.  On the other hand, under the same procedure, in a higher Landau level where stripes are  observed, they are  expected to survive and develop  gapless quantum Hall edge modes. 

The solvable model bears a certain resemblance to the wire construction \cite{kane02,kane14,kane17,kane17b}, in which abelian as well as non-abelian quantum Hall states are caused by interactions between Luttinger liquids describing stripes of interacting electrons. However, in the solvable model the one-dimensional systems are interlaced rather than separated in space and consists of neutral dipoles rather than charged electrons. The solvable model describing a gapless phase is obtained without interactions between the one-dimensional systems. 

\begin{acknowledgments}
I would like to thank Emil J. Bergholtz, and Thors Hans Hansson for helpful and stimulating discussions.
\end{acknowledgments}

\appendix

\section{Ground State Homogeneity} \label{gshomo}

We here show that the ground state in the exact solution is homogeneous. Consider the density operator ${\rho({\bf r})}$=$\sum_{m,n=0}^{N_s-1} \psi_m^*({\bf r})\psi_n({\bf r}) c^{\dagger}_m c_n$ and a general normalized $T_1$-eigenstate $\ket{\chi}=\sum_j\gamma_j \ket{\zeta_j}$ in ${\cal H}'_0$, where $\ket{\zeta_j}$ are base states $\prod_{i=1}^t(a_i1_{n-2}b_i0_{n-2})$ (which all have the same $T_1$-eigenvalue). Form the  $T_2^2$-eigenstate $\ket{\Psi}=N_e^{-1/2}\sum_{j=0}^{N_e-1}\ket{\chi_{2j}}$, where
$\ket{\chi_{2j}}=e^{i4j\pi K_2 / N_s}T_2^{2j}\ket{\chi}$.  From the fact that the Hilbert spaces ${\cal H}'_r$ are disjoint and that $\ket{\zeta_j}$ all have the same  $T_1$-eigenvalue it follows that
$\bra{\zeta_j}T_2^{2p}c^{\dagger}_m c_n T_2^{2q}\ket{\zeta_k} =\beta_m \delta_{pq}\delta_{jk}\delta_{mn}$, where $\beta_m =1, (0)$ if site $m$ is occupied, (unoccupied), in $T_2^{2q}\ket{\zeta_k}$. This gives 
\begin{eqnarray} \label{homo}
\bra{\Psi} c^{\dagger}_mc_n\ket{\Psi}= {\frac {1} {N_e}} \sum_j |\gamma_j|^2    {\frac {N_e} {2} } \delta_{mn} = {\frac 1 2} \delta_{mn} \ \ .
\end{eqnarray}
Here, the 1/2 on the inert sites comes from half of them being occupied, whereas the 1/2 on the $a_i,b_i$-sites comes from the exact solution ground state in $\tilde {\cal H}$ for which  $\expect{c^{\dagger}_m c_n}=\half \delta_{mn}$ \cite{bk3}.
Hence, the expectation value of the density operator in the state $\ket{\Psi}$ is $\bra{\Psi} \rho({\bf r})\ket{\Psi}=\sum_{m,n=0}^{N_s-1} \psi_m^*({\bf r})\psi_n({\bf r}) \bra{\Psi} c^{\dagger}_mc_n\ket{\Psi}=\half \sum_{m=0}^{N_s-1} |\psi_m({\bf r})|^2$. Up to the factor 1/2  this is the density of the filled Landau level which is homogeneous when $N_e, \, L_1 \rightarrow \infty$. 
\\
\\

\section{$V_{k0}$ Ground State Away From $L_1=L_2$} \label{Vk0 gs}

It is instructive to consider the $V_{k0}$ ground state for a  system where $L_1$, and hence the lattice constant,  is kept fixed and $N_e$, and hence $L_2$, varies.  We choose $L_1$ so that it equals $L_2$ for $N_e=16$, $L_1=\sqrt{4 \pi N_e}=8\sqrt{\pi}$. For $N_e=4j$, $j=1,2,\dots $, results  indicate that the $V_{k0}$ ground state is the regular state $(1_n0_n)_t$, where $\{n,t\}$ are,
\begin{eqnarray} \label{gsL1}
&&\{4, 1\}, \{4, 2\}, \{4, 3\}, \{4, 4\}, \{5, 4\}, \{4, 6\}, \{4, 7\}, \{4, 8\}, \nonumber \\
&& \{4, 9\}, \{5, 8\}, \{4, 11\}, \{4, 12\} , \dots \ \  .
\end{eqnarray}
For the first four, \ie for $N_e=4,8,12,16$, the result has been proven, whereas for larger $N_e$ it has  been verified that the state in Eq. \ref{gsL1} has the lowest $V_{k0}$ energy among all the possible regular partitionings $(1_n0_n)_t$. When $L_1$ is kept fixed, the distance between the one-particle states along the $y$-axis is unchanged and it is natural that the ground state is simply a repetition of the same unit cell when $N_e$ increases in multiples of $n=4$.  Exceptions to this are seen at $N_e=20,40$ where the ground states are $(1_50_5)_4$ and $(1_50_5)_8$ instead of  $(1_40_4)_5$ and $(1_40_4)_5$, the latter  are here however the next lowest regular partitionings and have energies very close to the lowest ones. $(1_50_5)_4$ and $(1_50_5)_8$ fit into the series based on $L_1=L_2$ at $N_e=5^2=25$.  

If instead $L_2$ is kept fixed and $L_1 $ varies as a function of $N_e=4j$,  then $\{n,t\} \rightarrow \{t,n\}$ in Eq. \ref{gsL1} gives the $V_{k0}$ ground states. Thus, interchanging $L_1$ and $L_2$ gives a conjugate $V_{k0}$ ground state:  
$(1_n0_n)_t \, \rightarrow \, (1_t0_t)_n$.  In the extreme limit of very small $L_1$ the ground state is the Tao-Thouless state $\{1,N_e\}=(10)_{N_e}$ with unit cell 10. Interchanging $L_1$ and $L_2$ gives the conjugate state $\{N_e,1\}=1_{N_e}0_{N_e}$, this is the $V_{k0}$ ground state since the minimum at $k=1$ dominates when $L_1 \rightarrow \infty$, see Fig. \ref{Vk0}.  However,  while the TT-state  $(10)_{N_e}$ is the ground state for the full Coulomb interaction when $L_1\rightarrow 0$, since $V_{km}, m \neq 0,$ vanish in this limit, the conjugate state is not, since $V_{km}, m \neq 0,$ are non-zero when $L_1 \rightarrow \infty$.

\end{document}